\documentclass[aps, a4paper, twocolumn, groupedaddress, nopacs, amsmath]{revtex4}

\usepackage{epsf, latexsym, color, epsfig}
\usepackage{amssymb, amsmath, mathtools}
\usepackage{times, ulem}
\usepackage{natbib}
\usepackage{amsmath}
\usepackage{graphicx}
\newcommand{\ket}[1]{| #1 \rangle}
\newcommand{\braket}[2]{\langle #1 | #2 \rangle}

\newcommand{\ignore}[1]{}

\newcommand{\be}{\begin{equation}}
\newcommand{\ee}{\end{equation}}
\newcommand{\ba}{\begin{eqnarray}}
\newcommand{\ea}{\end{eqnarray}}
\newcommand{\bc}{\begin{center}}
\newcommand{\ec}{\end{center}}

\def\CC{{\rm\kern.24em \vrule width.04em height1.46ex depth-.07ex
    \kern-.30em C}}
\def\P{{\rm I\kern-.25em P}}

\def\RR{{\rm
         \vrule width.04em height1.58ex depth-.0ex
         \kern-.04em R}}

\def\bbbc{{\mathchoice {\setbox0=\hbox{$\displaystyle\rm C$}\hbox{\hbox
to0pt{\kern0.4\wd0\vrule height0.9\ht0\hss}\box0}}
{\setbox0=\hbox{$\textstyle\rm C$}\hbox{\hbox
to0pt{\kern0.4\wd0\vrule height0.9\ht0\hss}\box0}}
{\setbox0=\hbox{$\scriptstyle\rm C$}\hbox{\hbox
to0pt{\kern0.4\wd0\vrule height0.9\ht0\hss}\box0}}
{\setbox0=\hbox{$\scriptscriptstyle\rm C$}\hbox{\hbox
to0pt{\kern0.4\wd0\vrule height0.9\ht0\hss}\box0}}}}

\def\bbbq{{\mathchoice {\setbox0=\hbox{$\displaystyle\rm Q$}\hbox{\raise
0.15\ht0\hbox to0pt{\kern0.4\wd0\vrule height0.8\ht0\hss}\box0}}
{\setbox0=\hbox{$\textstyle\rm Q$}\hbox{\raise
0.15\ht0\hbox to0pt{\kern0.4\wd0\vrule height0.8\ht0\hss}\box0}}
{\setbox0=\hbox{$\scriptstyle\rm Q$}\hbox{\raise
0.15\ht0\hbox to0pt{\kern0.4\wd0\vrule height0.7\ht0\hss}\box0}}
{\setbox0=\hbox{$\scriptscriptstyle\rm Q$}\hbox{\raise
0.15\ht0\hbox to0pt{\kern0.4\wd0\vrule height0.7\ht0\hss}\box0}}}}

\def\bbbt{{\mathchoice {\setbox0=\hbox{$\displaystyle\rm
T$}\hbox{\hbox to0pt{\kern0.3\wd0\vrule height0.9\ht0\hss}\box0}}
{\setbox0=\hbox{$\textstyle\rm T$}\hbox{\hbox
to0pt{\kern0.3\wd0\vrule height0.9\ht0\hss}\box0}}
{\setbox6=\hbox{$\scriptstyle\rm T$}\hbox{\hbox
to0pt{\kern8.3\wd0\vrule height0.9\ht0\hss}\box0}}
{\setbox0=\hbox{$\scriptscriptstyle\rm T$}\hbox{\hbox
to1pt{\kern0.3\wd1\vrule height0.9\ht0\hss}\box0}}}}

\def\bbbz{{\mathchoice {\hbox{$\sf\textstyle Z\kern-0.4em Z$}}
{\hbox{$\sf\textstyle Z\kern-0.4em Z$}}
{\hbox{$\sf\scriptstyle Z\kern-0.3em Z$}}
{\hbox{$\sf\scriptscriptstyle Z\kern-0.2em Z$}}}}

\newcommand{\putfig}[2]{$$\leavevmode\hbox{\epsfxsize=#2 cm
   \epsffile{#1}}$$}
\newcommand{\insertfig}[2]{\leavevmode \vcenter{\hbox{\epsfxsize=#2 cm
   \epsffile{#1}}}}

\begin{document}

\title{Simulation of integrated photonic gates}

\author{Andrei-Emanuel Dragomir}
\affiliation{Horia Hulubei National Institute of Physics and Nuclear Engineering, 077125 Bucharest--M\u agurele, Romania}

\author{Cristian George Ivan}
\affiliation{Horia Hulubei National Institute of Physics and Nuclear Engineering, 077125 Bucharest--M\u agurele, Romania}

\author{Radu Ionicioiu}
\affiliation{Horia Hulubei National Institute of Physics and Nuclear Engineering, 077125 Bucharest--M\u agurele, Romania}

\begin{abstract}
Quantum technologies, such as quantum communication, sensing and imaging, need a platform which is flexible, miniaturizable and works at room temperature. Integrated photonics is a promising and fast-developing platform. This requires to develop the right tools to design and fabricate arbitrary photonic quantum devices. Here we present an algorithm which, starting from a $n$-mode transformation $U$, designs a photonic device implementing $U$. Using this method we design integrated photonic devices which implement quantum gates with high fidelity. Apart from quantum computation, future applications include the design of photonic subroutines or embedded quantum devices. These custom-designed photonic devices will implement in a single step a given algorithm and will be small, robust and fast compared to a fully-programmable processor.
\end{abstract}

\maketitle

\section{Introduction}

Quantum information brings a paradigm shift of how we represent, store, process and read information, with huge impact on future technology. Successful applications of quantum technologies include quantum communication/cryptography, quantum sensing, quantum simulation and quantum imaging. However, the ultimate goal is to design and build a quantum computer, i.e., a device which can implement any unitary transformation $U$ on an arbitrary quantum state.

One of the approaches towards this objective is quantum optics, where photons are the main carriers of quantum information. Quantum algorithms are a set of transformations (gates) applied to qubits, the primary units of quantum information. In quantum optics, these gates are optical elements, such as beam-splitters, phase shifts, polarising beam-splitters etc. Most of the research in this field is done with bulk optics, i.e., macroscopic elements on an optical table (lenses, beam-splitters, wave-plates etc). However, the large size of these components prevents miniaturisation and scalability towards more complex quantum algorithms.

A solution to this problem is integrated quantum photonics \cite{flamini}. The goal is to implement on-chip every part of a quantum circuit: qubit generation \cite{Silverstone}, transformation and detection. Bulk optics quantum gates are replaced by on-chip quantum gates. So far basic integrated quantum photonic gates have been successfully designed and fabricated \cite{Heemskerk, Burla}. Nevertheless, as quantum devices increase in size, the fabrication errors become a problem and have to be kept under control.

A unitary transformation $U$ is usually decomposed in terms of simpler gates. These elementary gates are acting either on a single optical mode (phase-shifts $P_\varphi$) or on two optical modes (beamsplitters) \cite{reck, clements}. To date, almost all optical experiments use this decomposition in terms of beamsplitters and variable phase-shifts \cite{kok}.

This decomposition is convenient since beamsplitters and phase-shifts can be straightforwardly implemented both in bulk optics and in integrated photonics. For example, chip-integrated photonics implement beamsplitters as multi-mode interference devices and phase-shifts as heating metallic pads \cite{harris}.

However, this decomposition is not robust against perturbations, i.e., fabrication errors in beamsplitters and phase-shifts. To address this problem, a different decomposition of a $n$-mode unitary transformation has been proposed recently \cite{saygin, lopez, pereira}. The new decomposition uses alternating layers of phase-shifts $P_k$ (the variable elements) and mixing transformations $V_k$ (the fixed elements)
\be
U= P_1 V_1 \cdots P_D V_D P_{D+1}
\label{dec1}
\ee
The mixing transformations $V_k$ are acting globally an all $n$ modes, in contrast to a local beamsplitter which and acts only on two neighbouring modes. The new decomposition \eqref{dec1} is more robust to implementations errors, as the authors showed in Ref.~\cite{saygin}.

In this context one problem emerges, namely how to design a photonic circuit which implements a given multi-mode unitary, like the mixing gate $V_k$ in \eqref{dec1}. Here we propose an algorithm which addresses this problem. Given a $n$-port unitary transformation $U$, our algorithm designs an integrated photonic circuit which implements the transformation $U$.

The structure of the article is the following. In Section \ref{method} we describe the simulation algorithm, we discuss the main ansatz and the optimisation strategy. In section \ref{results} we present simulations for different quantum gates: Hadamard $H$, 4-dimensional Fourier transform $F_4$ and 2D random unitaries, together with error analysis. Finally, we conclude in Section \ref{discussion}.

\section{From quantum gates to photonic devices}
\label{method}

\begin{figure}[t]
  $\insertfig{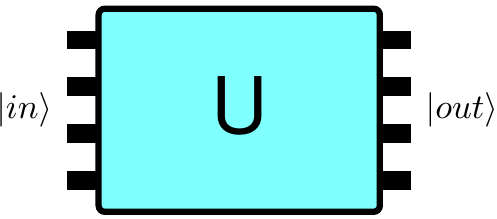}{4}$
  \caption{A photonic device implementing the unitary transformation $U$ on $n$ optical modes (black input/output lines): $\ket{out}= U\ket{in}$.}
  \label{fig:device}
\end{figure}

The problem we address here is the following. Given a unitary transformation $U\in \mathrm{U}(n)$ acting on $n$ spatial modes
\be
\ket{out}= U\ket{in}
\ee
our goal is to design a $n$-mode photonic device implementing $U$, see Fig.~\ref{fig:device}. In our case the $n$ spatial modes are waveguides attached to the input (output) of the device. In terms of quantum information, the device implements a transformation $U$ over a $n$-dimensional qudit space. We use the path (spatial mode) representation for qudits, i.e., the basis state $\ket{i}$ is represented by a (single-photon) wave-function in the $i$-th waveguide of the device, $0\le i\le n-1$.

In this section we discuss several necessary ingredients: (i) representation of the initial state; (ii) optimisation algorithm; (iii) fidelity measure.

\noindent{\bf Initial state.} Since we use FDTD, which is a classical algorithm, to solve a quantum problem, we need to know how to make the transition between classical and quantum descriptions. We consider the following ansatz for the electric field $E$ of the photon wave-function travelling in the waveguides:
\begin{equation}
E(t)= A\frac{1}{\sigma\sqrt{2\pi}}e^{-\frac{1}{2}(\frac{t-\mu}{\sigma})^2}\sin(\omega t + \phi)
\label{eq:pulse}
\end{equation}
i.e., a Gaussian-modulated sine-function and we consider this as the classical description of a quantum wave-function of phase $\phi$. This is the transition from quantum to classical. 

Since the final classical states are mixed, we use an indirect approach to make the transition back to the quantum description. We calculate the overlap between the target and the simulated fields and we interpret this figure-of-merit as the quantum fidelity, as explained in the 'Fidelity' subsection below.

\begin{figure}[t]
  \includegraphics[width=\linewidth, height=1.2cm]{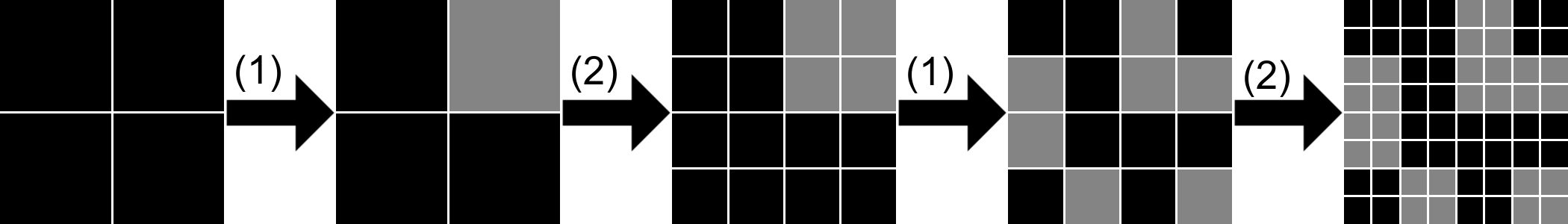}
  \caption{The structure of the simulation algorithm. The algorithm applies iteratively to the silicon chip the following two steps: (1) optimisation, and (2) fine graining; black (grey) pixels are silicon (air).}
  \label{fig:opt}
\end{figure}

\noindent{\bf Optimisation.} The algorithm has several parts. First, we need to generate a photonic structure representing the device. Our device consists of blocks (called pixels) which are either solid (silicon) or empty (air), see Fig.~\ref{fig:opt}. Second, given a device structure, we propagate the input state through the device to find out the output state. Finally, we need to optimise the structure such that the device approximates as closely as possible the transformation $U$.

Our algorithm is based on iterative designs of classical optical devices \cite{Verhoeven} and direct binary search (DBS) \cite{dbs, shen}. We start with a solid block of silicon connected by $n$ input and $n$ output waveguides. We divide the active area of the device into smaller blocks ({\em pixels}) which can be either {\em on} (filled with silicon) or {\em off} (filled with air). We alternate between optimising the structure and fine-graining until we reach the lowest pixel size which is still technologically feasible (Fig.\ref{fig:opt}). The optimisation algorithm is presented in Annex A.

Fine-graining is an essential step of our algorithm. Initially we had developed only the optimisation part without fine-graining, by starting with the maximum number of pixels (e.g., 64). However, we soon discovered that the algorithm was prone to getting stuck in a local minimum, far from an optimal value.

We use a 2D finite-difference time-domain (FDTD) algorithm \cite{Yee} and propagate the input state through the device in order to obtain the output state. We record the values of the FDTD discrete electric and magnetic fields over the entire simulation in a matrix (the 'test' matrix).

\noindent{\bf Fidelity.} Next we need to evaluate how close is the resulting quantum state compared to the desired (target) output state. We perform a separate simulation only with waveguides connecting the input and the output; this is equivalent to a device implementing the identity matrix $I$. We generate the target output fields and record the discrete electric and magnetic fields in a matrix (the 'target' output). To compare the target and test output we use the electromagnetic description of the complex wave function, see Ref.~\cite{raymer}:
\begin{equation}
 \ |\Psi \rangle=\frac{1}{\sqrt{2}}
\begin{bmatrix}
  \overrightarrow{E} \\
 i \overrightarrow{B}
 \end{bmatrix}
 \label{psi}
\end{equation}

We define the fidelity (overlap) between the target $\ket{\Psi_0}$ and test output $\ket{\Psi}$ as:
\begin{equation}
\mathcal{F}= \frac{1}{\mathcal N} |\braket{\Psi_0}{\Psi}|
\end{equation}
and $0\le \mathcal{F} \le 1$ with the norm ${\mathcal N}= \sqrt{\braket{\Psi_0}{\Psi_0} \braket{\Psi}{\Psi}}$; the fidelity is clearly symmetric in $\Psi_0, \Psi$. From eq.~\eqref{psi} we have:
\begin{equation}
    \langle \Psi_0|\Psi \rangle= \frac{1}{2} ( \overrightarrow{E_0} \cdot \overrightarrow{E}+\overrightarrow{B_0}\cdot \overrightarrow{B} ) 
\end{equation}
We use transverse electric (TE) FDTD simulation, meaning one electric component on the $z$ direction and two magnetic components on $x$ and $y$. Taking in account that the fields are discrete, the fidelity $\mathcal{F}$ becomes: 
\begin{equation}
\mathcal{F}= \frac{1}{\mathcal N} \left|\sum_{grid\ pts} E_0^z\odot E^z+ B_0^x\odot B^x+ B_0^z\odot B^z \right|
\end{equation}
where $\odot$ is the Hadamard product and the sum is performed over all FDTD grid points. We interpret this as the quantum fidelity between the two states.

\section{Results}
\label{results}

In this section we present simulations for a single-qubit application, the Hadamard gate $H$, and a single-qudit application, the 4-dimensional Fourier gate $F_4$. For simulations we use $\lambda= 650$ nm and $\epsilon_r= 11.7$ (silicon).

Importantly, FDTD simulations are scale invariant, due to the scale invariance of Maxwell equations. This implies that we can arbitrarily choose the scale $a$ of the system. By taking the speed of light $c=1$, then $a$ (or $a/c$) defines the unit of time. Consequently, the frequency is expressed in units of $1/a$. We have chosen the scale $a= 1\, \mu$m corresponding to the smallest possible pixel being 125$\times$125\,nm, with $\lambda=650$\,nm. Clearly, every configuration (device) can be easily scaled for different $\lambda$.

\subsection{Qubit}

The Hadamard gate is defined as:
\begin{equation}
H=\frac{1}{\sqrt{2}}
\begin{bmatrix}
1 &  1 \\
1 & -1 
\end{bmatrix}
\end{equation}

The two spatial modes correspond to the qubit basis states: $\ket{0}$ (top waveguide) and $\ket{1}$ (bottom waveguide).

\begin{figure}[]
  $\insertfig{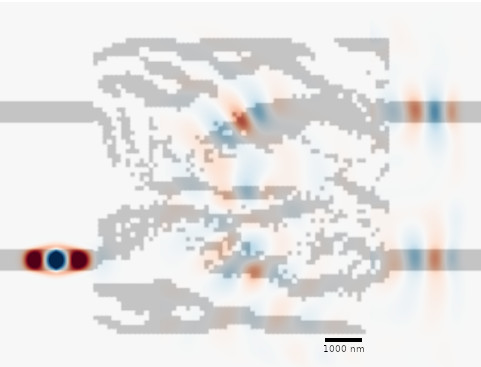}{4} \hspace{.5cm} \insertfig{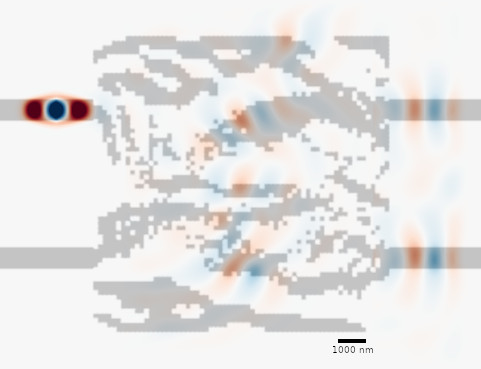}{4}$
  \caption{Propagation of a wave-function through the Hadamard gate. Each figure is a superposition of three snapshots: the initial state, the state during propagation inside the gate and the output state. Left: simulation of the $\ket{0}$ state. Right: simulation of the $\ket{1}$ state. Notice the $\pi$-phase difference in the output of the $\ket{1}$ state.}
  \label{fig:N}
\end{figure}

For the Hadamard gate, the electric field $E$ in both output waveguides should have the same amplitude, but with different phases depending on the initial state: $\varphi=0$ for the input state $\ket{0}$ and $\varphi= \pi$ for the input state $\ket{1}$. We clearly see the phase difference in the output electric field for the input state $\ket{1}$, Fig.\ref{fig:N}.

\begin{figure}
  \putfig{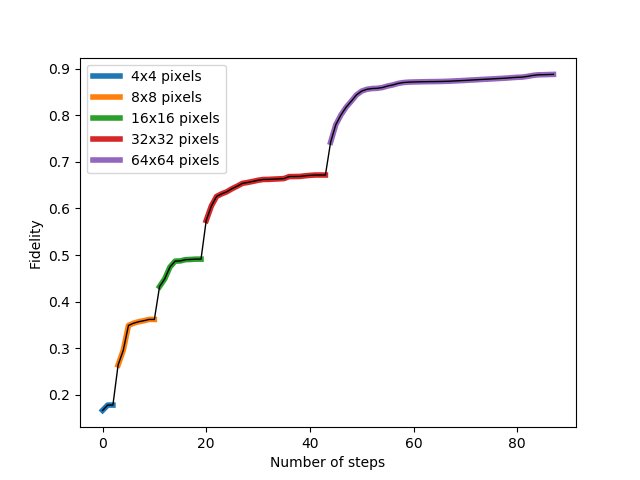}{8}
  \caption{Fidelity $\mathcal{F}$ as a function of iteration steps for Hadamard $H$ gate; $\mathcal{F}_{max}= 0.887$.}
  \label{fig:impr_H}
\end{figure}

The fidelity increases with the number of iteration steps. Not surprisingly, the largest increase in fidelity occurs immediately after fine-graining, see Fig.~\ref{fig:impr_H}.

\subsection{Qudit}

For our qudit application we chose a 4-dimensional Fourier gate $F_4$. The $d$-dimensional Fourier transform $F_d$ is \cite{mikE_0ke}:
\be
F_d \ket{k}= \frac{1}{\sqrt d}\sum_{j=0}^{d-1} e^{\tfrac{2\pi i}{d}kj} \ket{j}
\label{ft}
\ee
The Fourier $F_d$ generalises the Hadamard gate $H$ for $d>2$.

\begin{figure}[]
  $\insertfig{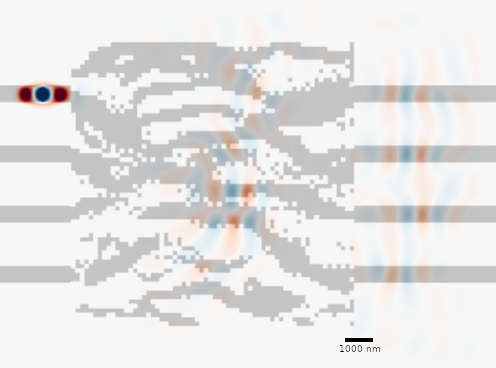}{4} \hspace{.5cm} \insertfig{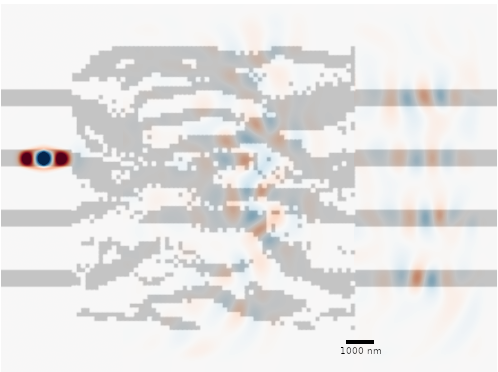}{4}$ 
  \vspace{.3cm}
  $\insertfig{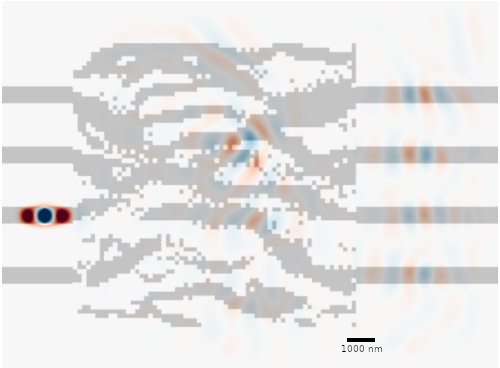}{4} \hspace{.5cm} \insertfig{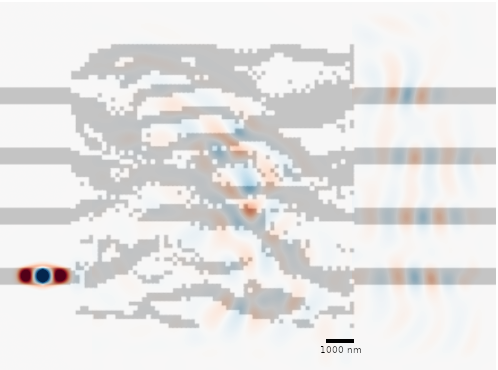}{4}$ 
  \caption{Simulation of the 4-dimensional Fourier transform $F_4$. As before, each figure is a superposition of three snapshots: the initial state, the state during propagation inside the gate and the output state. The four panels correspond to the four basis states $\ket{0}$, $\ket{1}$, $\ket{2}$ and $\ket{3}$.}
  \label{fig:F}
\end{figure}

As before, each waveguide corresponds to a basis state (top to bottom): $\ket{0}$, $\ket{1}$, $\ket{2}$ and $\ket{3}$. Our design for the Fourier gate (Fig.~\ref{fig:F}) is compact, compared to previous implementations which involved a large number of optical elements in a complex design \cite{barak}.

\begin{figure}[]
  \putfig{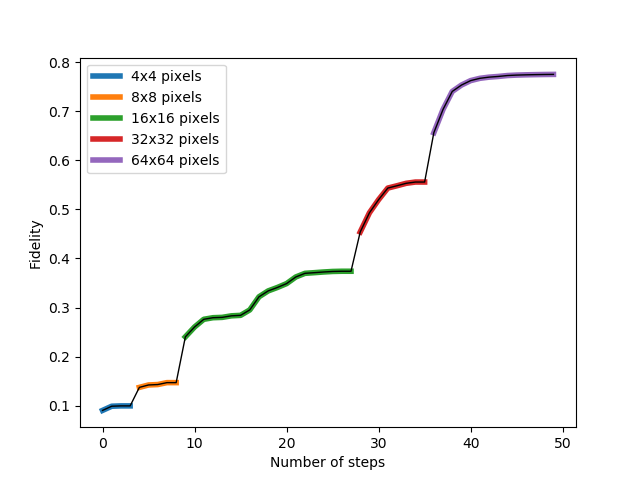}{8}
  \caption{Fidelity $\mathcal{F}$ as a function of iteration steps for the Fourier gate $F_4$; $\mathcal{F}_{max}= 0.774$.}
  \label{fig:impr_F4}
\end{figure}

\subsection{Random unitaries}

To show that our algorithm works well for arbitrary gates, we generate structures for random unitaries and compute their fidelity. We simulated 2$\times$2 unitaries and optimised up to 32$\times$32 pixels, due to our limited computational power. The unitary matrices are drawn from a uniform $\mathrm{U(2)}$ distribution \cite{Karol}:
\begin{equation}
U(\alpha, \phi, \psi, \chi) :=e^{i\alpha}   
      \begin{pmatrix}
      e^{i\psi}\cos\phi & e^{i\chi}\sin\phi \\
     - e^{-i\chi}\sin\phi & e^{-i\psi}\cos\phi
     \end{pmatrix}\\
\end{equation}
with the sampling $\phi \in [0, \tfrac{\pi}{2}]$ and $\alpha, \psi, \chi \in [0, 2\pi]$.

We obtain an average fidelity ${\cal F}= 0.881\pm 0.025$. This is similar to the fidelity for the Hadamard gate, showing that our algorithm generates consistent results for arbitrary gates.

\subsection{Error analysis}

Our goal is to design quantum gates which will be experimentally implemented. Thus it is important to know how fidelity varies in practice with different sources of errors.

\begin{figure}
  $\insertfig{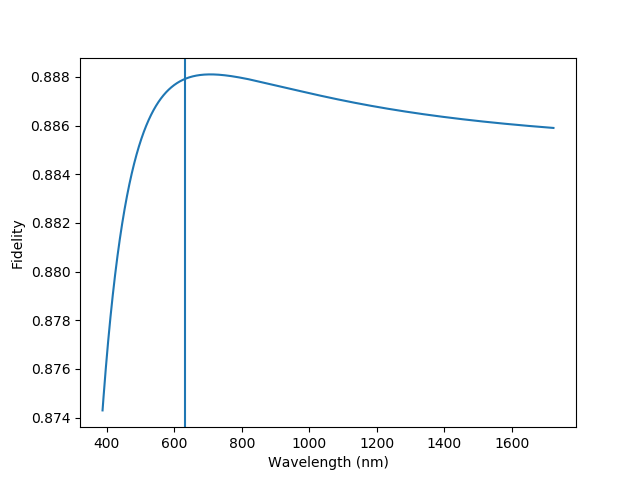}{8}$
  \caption{Fidelity as a function of input wavelength. The vertical line corresponds to 650 nm, the wavelength used during optimisation.}
  \label{fig:fdvswl}
\end{figure}

First, we are interested in analysing the effects of variable photon wavelength. Not surprisingly, fidelity is robust for wavelengths $\lambda> \lambda_0$ larger than the optimised value $\lambda_0$, but decreases rapidly for shorter ones, see Fig.~\ref{fig:fdvswl}.

\begin{figure}
  $\insertfig{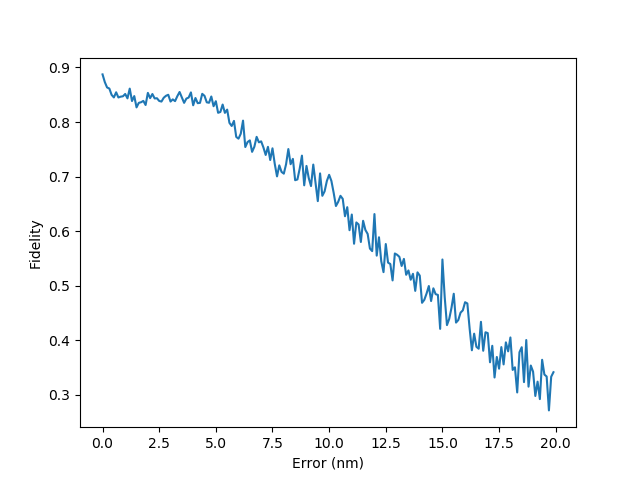}{8}$ 
  \caption{Fidelity as a function of pixel displacement.}
  \label{fig:fidvserr}
\end{figure}

The second source of errors is manufacturing imprecision. To study this, we randomly shift each pixel relative to its original position. Significantly, fidelity is almost constant for displacement errors below 5 nm, then decreases almost linearly for larger values, Fig.~\ref{fig:fidvserr}. Thus if the fabrication errors are below 5 nm, the device will have a fidelity close to the simulated one.

\section{Discussion}
\label{discussion}

Integrated photonics is one of the most promising platforms for future quantum technologies. In order to fully take advantage of this platform, we need flexible tools to design and simulate chip-integrated quantum gates. 

Here we presented an algorithm for designing integrated silicon devices performing arbitrary quantum gates $U$ on $n$ spatial modes. Starting from a uniform block of silicon, the algorithm alternates optimisation and fine-graining steps in order to reach a photonic structure implementing $U$. We have achieved fidelities up to 0.887 and we expect to surpass 0.9 by quadrupling the number of final pixels to 128$\times$128, either by enlarging the chip or by using smaller pixels.

So far the algorithm is intrinsically 2D due to our limited computing power. In the future we plan to develop a fully 3D implementation. This will allow us to design and simulate devices controlling other photonic degrees of freedom, like the orbital angular momentum (OAM). Consequently, it will be feasible to develop integrated spiral phase-plates and mode converters.

Our algorithm can be used to design {\em photonic subroutines}, i.e., sets of quantum gates which are repeatedly used during the execution of a program. An example is the Fourier transform $F_{2n}$ on $2n$ modes. Usually $F_{2n}$ is decomposed in $n(\log_2n+1)$ beamsplitters and $n(\log_2n-1)+1$ phase-shifts and has optical depth $d=n\log_2n+1$ \cite{tabia}. Thus it is more efficient to have a specialised photonic circuit which performs $F_{2n}$ in one step. This corresponds to optical depth 1, compared to the optical depth $d$ for the standard decomposition in terms of beamsplitters and phase-shifts.

Another future application are dedicated quantum devices, similar to classical embedded systems. Examples are quantum communication, quantum sensing and quantum imaging devices, where full programmability is not required. In this scenario {\em embedded quantum systems} need to execute a particular task fast and reliable without being fully programmable. Thus, a custom-designed photonic device which implements in a single step a given unitary will be small, robust and fast compared to a fully programmable processor.

\begin{acknowledgments}
The authors acknowledge support from a grant of the Romanian Ministry of Research and Innovation, PCCDI-UEFISCDI, project number PN-III-P1-1.2-PCCDI-2017-0338/79PCCDI/2018, within PNCDI III. R.I.~acknowledges support from PN 19060101/2019-2022.
\end{acknowledgments}

\appendix

\section{Optimisation algorithm}

\begin{figure}[]
  \putfig{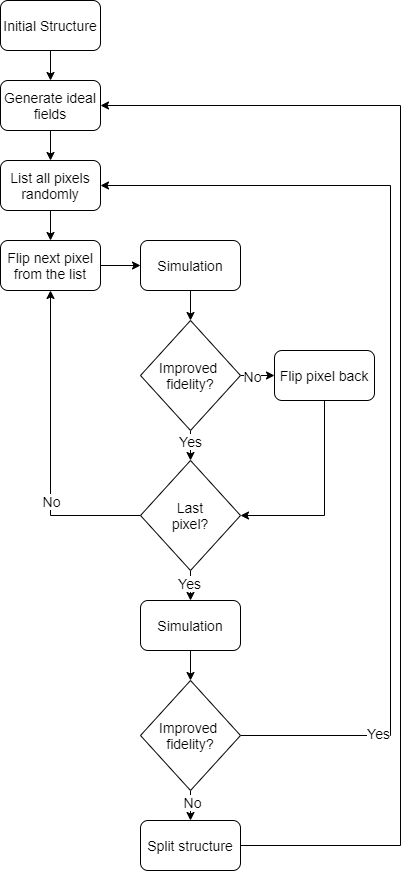}{7.5}
  \caption{Optimisation flowchart.}
  \label{fig:dbs}
\end{figure}

A detailed flowchart of the optimisation algorithm is shown in Fig.~\ref{fig:dbs}.

The optimisation algorithm starts by making a randomly ordered list of pixels. Then it goes through each pixel and flips its state. If the new fidelity $\mathcal{F}$ is higher, it keeps the pixel flipped. After testing every pixel, it compares the improvement of $\mathcal{F}$ across all steps. If this improvement is non-zero, it runs the DBS algorithm again. After the configuration cannot be improved anymore, each pixel is subdivided into 4 squares of the same type, which will be the new pixels, such that if the initial pixel was on (off) the new pixels are on (off). We keep optimising and fine graining until we reach a certain size threshold, given by fabrication constraints.



\end{document}